\documentclass[twocolumn,showpacs,preprintnumbers,amsmath,amssymb]{revtex4}
\usepackage{amssymb}
\usepackage{graphicx}
\usepackage{dcolumn}
\usepackage{bm}
\usepackage{textcomp}

\newcommand\ket[1]{\ensuremath{|#1\rangle}}
\newcommand\bra[1]{\ensuremath{\langle#1|}}

\newcommand\oprod[2]{\ensuremath{|#1\rangle\langle#2|}}

\newcommand\mean[1]{\ensuremath{\langle #1 \rangle}}

\begin{document}
\title{Sending-or-not twin-field quantum key distribution in practice}
\author{Zong-Wen Yu$^{1,2}$, Xiao-Long Hu$^{1,3}$, Cong Jiang$^{1,3}$, Hai Xu$^{1,3}$, Xiang-Bin Wang$^{1,3,4\footnote{Email Address: xbwang@mail.tsinghua.edu.cn}\footnote{Also a member of Center for Atomic and Molecular Nanosciences at Tsinghua University}}$}

\affiliation{ \centerline{$^{1}$State Key Laboratory of Low
Dimensional Quantum Physics, Tsinghua University, Beijing 100084,
People's Republic of China}\centerline{$^{2}$Data Communication Science and Technology Research Institute, Beijing 100191, People's Republic of China}\centerline{$^{3}$ Synergetic Innovation Center of Quantum Information and Quantum Physics, University of Science and Technology of China}\centerline{  Hefei, Anhui 230026, People's Republic of China
 }\centerline{$^{4}$ Shandong
Academy of Information and Communication Technology, Jinan 250101,
People's Republic of China}}

\begin{abstract}
We present results of practical sending-or-not quantum key distribution. In real-life implementations, we need consider the following three requirements, a few different intensities rather than infinite number of different intensities, a phase slice of appropriate size rather than infinitely small size and the statistical fluctuations. We first show the decoy-state method with only a few different intensities and a phase slice of appropriate size. We then give a statistical fluctuation analysis for the decoy-state method. Numerical simulation shows that, the performance of our method is comparable to the asymptotic case for which the key size is large enough. Our results show that practical implementations of the sending-or-not quantum key distribution can be both secure and efficient.
\end{abstract}


\maketitle


\section{Introduction}\label{SecIntro}
Quantum key distribution (QKD) allows two parties, Alice and Bob, to share unconditional secret keys based on the laws of quantum physics~\cite{BB84,GRTZ02}, even in the presence of an eavesdropper, Eve. However, in real-life implementations of QKD, it's practical security is still questionable due to the device imperfections, such as the imperfect source and detectors. Fortunately, by using the decoy-state method~\cite{ILM,H03,wang05,wang06,LMC05,AYKI,haya,peng,wangyang,rep,njp}, it has been shown that the unconditional security of QKD can still be assured with an imperfect single-photon source~\cite{PNS1,PNS}. To avoid the detector side channel attacks, the measurement-device-independent QKD (MDI-QKD) was proposed~\cite{curty1,wang10}. The decoy-state MDI-QKD can remove all detector side-channel attack with imperfect single-photon sources.

With the developments~\cite{ILM,H03,wang05,wang06,LMC05,AYKI,haya,peng,wangyang,rep,njp,PNS1,PNS,curty1,wang10,MDIExps,MDI404km} in both theory and experiment, it is more and more hoped to extensively applied in practice, though there are barriers for so. Among them, the transmission loss of photons for long distance QKD has become the major obstacle in practical implementations. Very recently, with the technology of long distance single-photon interference, the twin-field quantum key distribution (TF-QKD) has been presented~\cite{nature18}. The original TF-QKD does not offer the information-theoretic security. By switching between a Test mode and a Code mode, the TF-QKD* protocol with information-theoretic security has been shown~\cite{star}. A more efficient protocol for TF-QKD with sending or not sending the coherent state has been given in~\cite{wxb}. In the sending-or-not protocol~\cite{wxb}, Alice and Bob do not take post selection for the bits in $Z$ basis (signal pulses) and hence the traditional calculation formulas directly apply. 

In Ref.~\cite{wxb}, a security analysis is provided for sending-or-not QKD asymptotically. In the asymptotic case, the number of different intensities is infinite, the phase slice is infinitely small and the key size is large enough. However, in any real implementations, these requirements can no be fulfilled. In practice, we need consider the situations with a few different intensities rather than infinite number of different intensities, a phase slice of appropriate size rather than infinitely small size and the statistical fluctuations. In this paper, we proceed further and analysis the performance of the sending-or-not QKD under the above real-life assumptions.

First, we need reveal the decoy-state method with only a few different intensities and a phase slice of appropriate size to estimate the lower bound of the yield and the upper bound of the phase-flip error rate for the single-photon state. Furthermore, we also need to consider the statistical fluctuations. In order to make the utmost decrease the effect of statistical fluctuations, the instances for basis unmatched are also used to estimate the lower bound of the yield for the single-photon state.

The rest of this paper is organized as follows. In Sec.~\ref{sec:Protocol}, we shall first review the four-intensity decoy-state method for sending-or-not QKD, and then perform a statistical fluctuation analysis on it in Sec.~\ref{sec:Sta}. We then present the numerical simulation results in Sec.~\ref{sec:Numerical}. The article is ended in Sec.~\ref{sec:Conclusion} with a concluding remark.

\section{The decoy-state method with a few different intensities and a phase slice of appropriate size}\label{sec:Protocol}
In the four-intensity decoy-state sending-or-not protocol, Alice and Bob randomly choose the $X$-window (decoy pulses) and $Z$-window (signal pulses) to send or not to send a phase-randomized coherent pulse to an untrusted party, Charlie, who is expected to perform interference measurement. When one and only one of the two detectors clicks, Charlie obtains an effective event. Then Alice and Bob can distill the final secret key with post-selection and post-processing. The protocol is detailed below.

\begin{itemize}
  \item[1.] Alice and Bob repeat Steps 2-3, $N$ times. All the public announcements by the legitimate users Alice and Bob are done over an authenticated channel.
  \item[2.] Alice and Bob randomly choose $X$-window and $Z$-window with probabilities $p_X$ and $1-p_X$ respectively. In $X$-window, both Alice and Bob prepare and send the decoy pulses. Explicitly they randomly choose three sources $\rho_{\alpha_i}$ with probability $p_i$ for $i=0,1,2$, where $\rho_{\alpha_0}=\oprod{0}{0}$ is the vacuum source, $\rho_{\alpha_1}$ and $\rho_{\alpha_2}$ are two coherent sources with intensity $\mu_1$ and $\mu_2$ ($\mu_1<\mu_2$) respectively. In $Z$-window, Alice (Bob) randomly prepares and sends the coherent state $\rho_{\alpha_z}$ with probability $p_z$ and sends nothing else.
  \item[3.] Charlie measures the incoming signals and records which detector clicks. When the quantum communication is over, he publicly announces all the information about the detection event. The situation when one and only one detector (detector 0 or detector 1) makes a count is denoted as an effective event. Alice and Bob collect all the data with effective events and discard all the others.
  \item[4.] Alice and Bob announce the basis information ($X$-window or $Z$-window) firstly. Then they announce the bit values and phase information corresponding to the effective events when Alice or Bob choose $X$-window. With these information, Alice and Bob obtain the observable $N_{jk}(j,k=0,1,2,z)$ being the number of instances when Alice and Bob send state $\rho_{\alpha_j}$ and $\rho_{\alpha_k}$ respectively. Correspondingly, the lowercases $n_{jk}$ are used to denote the number of effective events. The yields can be defined as $S_{jk}=n_{jk}/N_{jk}$. Explicitly, we have
      \begin{eqnarray}
        N_{00}&=&p_0^2 N_X + 2 p_0(1-p_z) N_{XZ}, \nonumber \\
        N_{01}&=&N_{10}=p_0 p_1 N_X + (1-p_z)p_1 N_{XZ}, \label{eq:DefN}\\
        N_{02}&=&N_{20}=p_0 p_2 N_X + (1-p_z)p_2 N_{XZ}, \nonumber
      \end{eqnarray}
      where $p_0=1-p_1-p_2$ is the probability to send a vacuum pulse in $X$-window, $N_X=p_X^2 N$ is the number of instances when both Alice and Bob choose $X$-window and $N_{XZ}=p_X (1-p_X) N$ is the number of instances when Alice chooses $X$-window and Bob chooses $Z$-window.
  \item[5.] Define two sets $C_{\Delta^{+}}$ and $C_{\Delta^{-}}$ that contain the instances when both Alice and Bob send $\rho_1$ in $X$-window with the phase information $\theta_A$ and $\theta_B$ falling into the slice $|\theta_A-\theta_B|\leq \Delta/2$ and $|\theta_A-\theta_B-\pi|\leq \Delta/2$ respectively. The number of instances in $C_{\Delta^{\pm}}$ are $N_{11}^{\Delta^{\pm}}=\frac{\Delta}{2\pi} N_{11}$. The number of effective events corresponding to $C_{\Delta^{\pm}}$ are denoted by $n_{11}^{\Delta_0^{\pm}}$ and $n_{11}^{\Delta_1^{\pm}}$ for detector 0 and detector 1 respectively.
  \item[6.] With these observable in Step.5, Alice and Bob can estimate the lower bound of $s_1$ and the upper bound of $e_1^{ph}$ by using the decoy-state methods shown below. Then the post-processing can be performed and the final key length can be calculated with the following formula
      \begin{equation}\label{eq:KeyRateN}
        N_{f}= n_1 [1-H(e_1^{ph})]-f n_t H(E_Z),
      \end{equation}
      where $N_f$ is the number of final bits, $n_1$ is the number of effective events caused by single-photon states in $Z$-basis when Alice decides sending while Bob decides not sending or Alice decides not sending while Bob decides sending, $e_1^{ph}$ is the phase-flip error rate for instances of $n_1$, $n_t$ is the number of effective events when both Alice and Bob choose $Z$-window and  $E_Z$  is the corresponding bit-flip error rate.
\end{itemize}

In the above protocol, Alice and Bob prepare and send the coherent pulses with randomized phase. The traditional formulas of decoy-state method can be applied directly. The coherent state whose phase is selected uniformly at random can be regard as a mixture of photon number states
\begin{equation}\label{eq:DefCS}
  \rho_{\alpha_j}=e^{-\mu_j}\sum_{n=0}^{\infty} \frac{\mu_j^n}{n!}\oprod{n}{n}, \quad (j=0,1,2,z)
\end{equation}
where $\mu_j=|\alpha_j|^2$ is the intensity of the coherent state $\ket{\alpha_j}$. Then the state when Alice decides not sending and Bob decides to send $\rho_{\alpha_k}$ is $\rho_{\alpha_0\alpha_k}=e^{-\mu_k}\sum_{n=0}^{\infty} \frac{\mu_k^n}{n!}\oprod{0n}{0n}$. With these convex forms, the lower bound of the yield of the state $\rho_{z_0}=\oprod{01}{01}$ can be written into the following form
\begin{equation}\label{eq:sz0L}
  s_{z_0}\geq s_{z_0}^{L}=\frac{\mu_2^2 e^{\mu_1}S_{01} -\mu_1^2 e^{\mu_2} S_{02}-(\mu_2^2-\mu_1^2)S_{00}}{\mu_1 \mu_2 (\mu_2-\mu_1)},
\end{equation}
where the observable $S_{0k}$ are the yield of the sources $\rho_{0k}$ for $k=1,2$, $S_{00}$ is the yield when both Alice and Bob send the vacuum state. Similarly, the lower bound of the yield of the state $\rho_{z_1}=\oprod{10}{10}$ can be written as
\begin{equation}\label{eq:sz1L}
  s_{z_1}\geq s_{z_1}^{L}=\frac{\mu_2^2 e^{\mu_1}S_{10} -\mu_1^2 e^{\mu_2} S_{20}-(\mu_2^2-\mu_1^2)S_{00}}{\mu_1 \mu_2 (\mu_2-\mu_1)},
\end{equation}
where the observable $S_{j0}$ are the yield of the sources when Alice sends the coherent state $\rho_j$ and Bob sends the vacuum state for $j=1,2$. With Eq.(\ref{eq:sz0L}) and Eq.(\ref{eq:sz1L}), the lower bound of the yield of single-photon state in $Z$-basis, i.e., the state $\rho_1^Z=\frac{1}{2}(\rho_{z_0}+\rho_{z_1})$, has the following form
\begin{equation}\label{eq:s1ZL}
  s_1^Z\geq \underline{s}_1^{Z}=\frac{1}{2}(s_{z_0}^{L}+s_{z_1}^{L}).
\end{equation}

{\textbf{Note:}} Replacing the source $\rho_2$ used in Eqs.(\ref{eq:sz0L}-\ref{eq:s1ZL}) with the source $\rho_z$, we obtain the other lower bound of $s_1^Z$. With this replacement, source $\rho_2$ is not used actually, then the four-intensity decoy-state method can be simplified to a three-intensity decoy-state method by taking $p_2=0$. On the one hand, the three-intensity decoy-state method can be carried out easily in experiment. On the other hand, if we are not attention to the limit security distance but the usability key rate practically (such as $10^{-6}$ per-pulse), the key rate of the three-intensity decoy-state method is only a little lower than (less than one percent for the cases discussed in the numerical simulation) the results for the four-intensity decoy-state method.

In the rest of this section, we show the formula to estimate the upper bound of $e_1^{ph}$ in Eq.(\ref{eq:KeyRateN}) with the observable. The state of pulse pair when Alice sends the coherent state $\ket{\alpha_1^{A}=\sqrt{\mu_1}e^{i\theta_A}}$ and Bob sends the coherent state $\ket{\alpha_1^B=\sqrt{\mu_1}e^{i\theta_B}}$ is
\begin{equation}\label{eq:alpha1AB}
  \ket{\alpha_1^A} \ket{\alpha_1^B}=e^{-\mu_1}\sum_{n=0}^{\infty}\frac{(\sqrt{2\mu_1}e^{i\theta_B})^n}{\sqrt{n!}}\ket{\psi_{n}^{\delta^+}}.
\end{equation}
Similarly, we also have
\begin{equation}\label{eq:alpha1ABp}
  \ket{\alpha_1^A} \ket{-\alpha_1^B}=e^{-\mu_1}\sum_{n=0}^{\infty}\frac{(-\sqrt{2\mu_1}e^{i\theta_B})^n}{\sqrt{n!}}\ket{\psi_{n}^{\delta^-}}.
\end{equation}
In Eq.(\ref{eq:alpha1AB}) and Eq.(\ref{eq:alpha1ABp}), the $k$-photon twin-field state $\ket{\psi_n^{\delta^+}}$ is defined as follows
\begin{eqnarray}\label{eq:defphi}
  \ket{\psi_n^{\delta^{+}}}&=& \frac{1}{\sqrt{2^n}}\sum_{m=0}^{n}\frac{\sqrt{n!}e^{im\delta}}{\sqrt{m!(n-m)!}}\ket{m}\ket{n-m}, \label{eq:defphiP} \\
  \ket{\psi_n^{\delta^{-}}}&=& \frac{1}{\sqrt{2^n}}\sum_{m=0}^{n} \frac{(-1)^m\sqrt{n!}e^{im\delta}}{\sqrt{m!(n-m)!}}\ket{m}\ket{n-m}, \label{eq:defphiM}
\end{eqnarray}
where $\delta=\theta_A-\theta_B$. For the state in set $C_{\Delta^{+}}$, the phase is selected uniformly at random in the slice with $|\theta_A-\theta_B|\leq \Delta/2$. Equivalently, in set $C_{\Delta^{+}}$, the phase $\theta_B$ chosen by Bob in $\ket{\alpha_1^A}\ket{\alpha_1^B}$ can be regarded as uniformly distributed in $[0,2\pi)$ and the phase $\theta_A$ chosen by Alice satisfies the condition $|\delta|\leq \Delta/2$. For any fixed value $\delta$, we have
\begin{eqnarray}\label{eq:rhodeltaP}
  \rho_{\delta^{+}}&=&\frac{1}{2\pi}\int_0^{2\pi} \ket{\alpha_1^A}\ket{\alpha_1^B} \bra{\alpha_1^A}\bra{\alpha_1^B}d\theta_2 \nonumber \\
  &=& e^{-2\mu_1}\sum_{n=0}^{\infty} \frac{(2\mu_1)^{n}}{n!}\oprod{\psi_{n}^{\delta^+}}{\psi_{n}^{\delta^+}}.
\end{eqnarray}
Similarly, we also have
\begin{eqnarray}\label{eq:rhodeltaP}
  \rho_{\delta^{-}}&=&\frac{1}{2\pi}\int_0^{2\pi} \ket{\alpha_1^A}\ket{-\alpha_1^B} \bra{\alpha_1^A}\bra{-\alpha_1^B}d\theta_2 \nonumber \\
  &=& e^{-2\mu_1}\sum_{n=0}^{\infty} \frac{(2\mu_1)^{n}}{n!}\oprod{\psi_{n}^{\delta^-}}{\psi_{n}^{\delta^-}}.
\end{eqnarray}
Considering the single-photon twin-field states in $C_{\Delta}=C_{\Delta^{+}}\cup C_{\Delta^{-}}$ for a fixed $\delta$, we have
\begin{equation}\label{eq:Sdelta}
  \rho_1^{\delta} =\frac{1}{2}(\oprod{\psi_1^{\delta^+}}{\psi_1^{\delta^+}} + \oprod{\psi_1^{\delta^-}}{\psi_1^{\delta^-}})= \rho_1^{Z}.
\end{equation}
So we know that the single-photon states in set $C_{\Delta}$ and in $Z$-basis have the same density matrices. The probability to emit a single-photon pulse from $C_{\Delta}$ is $q_1=2\mu_1 e^{-2\mu_1}$. With this relations, we know that the bit-flip error rate of single-photon state in set $C_{\Delta}$ is equal to the phase-flip error rate $e_1^{ph}$ asymptotically. The bit-flip error rate for all instances in set $C_{\Delta}$ can be calculated with the observable as follows
\begin{equation}\label{eq:ErrorDelta}
  T_{\Delta}=\frac{1}{2}(T_{\Delta^+}+T_{\Delta^-})=\frac{1}{2}\left(\frac{n_{11}^{\Delta_1^{+}}}{N_{11}^{\Delta^+}} +\frac{n_{11}^{\Delta_0^{-}}}{N_{11}^{\Delta^-}}\right).
\end{equation}
Attribute all the error to the single-photon state and the vacuum state, the upper bound of phase-flip error rate $e_1^{ph}$ is
\begin{equation}\label{eq:e1phU}
  e_1^{ph}\leq \overline{e}_{1}^{ph}=\frac{T_{\Delta} -1/2e^{-2\mu_1} S_{00}}{2\mu_1 e^{-2\mu_1} \underline{s}_1^Z},
\end{equation}
where $\underline{s}_1^Z$ is the lower bound of $s_1^{Z}$ given in Eq.(\ref{eq:s1ZL}). Then the final key rate of per pulse can be calculated with
\begin{eqnarray}\label{eq:KeyRate}
  R&=&(1-p_X)^2\{2p_z(1-p_z) \nonumber \\
   & & a_1 s_1[1-H(e_1^{ph})]-f S_Z H(E_Z)\},
\end{eqnarray}
where $R$ is the final key rate, $a_1=\mu_z e^{-\mu_z}$ is the probability to emit a single-photon state from source $\rho_z$, $s_1$ is the yield of the single-photon state in $Z$-window when one party from Alice and Bob decides to send a signal states, $e_1^{ph}$ is the phase-flip error rate for those instance of $s_1$, $S_Z$ and  $E_Z$  are the yield and bit-flip error rate for instances when both Alice and Bob choose $Z$-window.

\section{Statistical fluctuation analysis}\label{sec:Sta}
In the real protocol, in order to extract the secure final key, we have to consider the effect of statistical fluctuations. To obtain the lower bound value for $s_1$ and the upper bound value for $e_1^{ph}$ in the real protocol with finite $N$, one can implement the idea of Ref.~\cite{njp}, i.e., treating the averaged yield. Accordingly, define $\mean{S}$ as the mean value of yield $S$. Note that even though $S_{jk}(j,k=0,1,2,z)$ are known values directly observed in the experiment, the mean values $\mean{S_{jk}}$ are not. However, given the observed values $S_{jk}$ and the corresponding number of pulse pairs, the confidence lower and upper limits of $\mean{S_{jk}}$ can be calculated.

In order to obtain a tighter lower bound of $\mean{s_1^Z}$, we need introduce the following two yields
\begin{eqnarray}
  S_1=\frac{1}{2} (S_{01}+S_{10}) =\frac{n_{01}}{2N_{01}} +\frac{n_{10}}{2N_{10}}, \label{eq:DefS1} \\
  S_2=\frac{1}{2} (S_{02}+S_{20}) =\frac{n_{02}}{2N_{02}} +\frac{n_{20}}{2N_{20}}, \label{eq:DefS2} 
\end{eqnarray}
Replacing the observed yields with their mean values in Eq.(\ref{eq:s1ZL}) and Eq.(\ref{eq:e1phU}) we can formulate the lower bound of $\mean{s_1^Z}$ and the upper bound of $\mean{e_1^{ph}}$ respectively. Explicitly, we have
\begin{equation}\label{eq:s1LM}
  \mean{s_1^Z}\geq \mean{\underline{s}_1^Z}=\frac{\mu_2^2 e^{\mu_1}\underline{S}_{1} -\mu_1^2 e^{\mu_2} \overline{S}_{2}-(\mu_2^2-\mu_1^2)\overline{S}_{00}}{\mu_1 \mu_2 (\mu_2-\mu_1)},
\end{equation}
and
\begin{equation}\label{eq:e1phUM}
  \mean{e_1^{ph}}\leq \mean{\overline{e}_1^{ph}}=\frac{\overline{T}_{\Delta} -1/2e^{-2\mu_1} \underline{S}_{00}}{2\mu_1 e^{-2\mu_1} \mean{\underline{s}_1^Z}},
\end{equation}
with
\begin{equation}\label{eq:LUofSy01}
  \underline{S}_{k}={S}_{k}/(1+\delta_{k}), \quad \overline{S}_{k}={S}_{k}/(1-\delta_{k}^{\prime}).
\end{equation}
By using the multiplicative form of the Chernoff bound~\cite{wang10}, with a fixed failure probability $\epsilon$, we can give an interval of $\mean{S_{k}}$ with the observable $S_{k}$, $[\underline{S}_{k}, \overline{S}_{k}]$, which can bound the value of $\mean{S_{k}}$ with a probability of at least $1-\epsilon$.

With the mean values $\mean{\underline{s}_1^Z}$ and $\mean{\overline{e}_1^{ph}}$ defined in Eq.(\ref{eq:s1LM}) and Eq.(\ref{eq:e1phUM}), the lower bound of the yield $\underline{s}_1$ and the upper bound of the phase-flip error rata $\overline{e}_1^{ph}$ corresponding to $s_1$ in Eq.(\ref{eq:KeyRate}) can be estimated by
\begin{equation}\label{eq:s1cL}
  \underline{s}_1=\mean{\underline{s}_1^Z}(1-\delta_1^c), \quad \overline{e}_1^{ph}=\mean{\overline{e}_1^{ph}}(1+\delta_1^{\prime c}).
\end{equation}

With the lower bound of $s_1$ and the upper bound of $e_{1}^{ph}$ in Eq.(\ref{eq:s1cL}), the final key rate can be calculated with Eq.(\ref{eq:KeyRate}).

{\textbf{Note:}} In Eq.(\ref{eq:s1LM}), by using the method shown in~\cite{wang10}, we can treat the observable $S_{2}$ and $S_{00}$ jointly when we considering the effect of statistical fluctuation. Then we obtain a tighter lower bound of $\mean{s_{1}^{L}}$. Furthermore, $S_{00}$ is the common variable in both the lower bound of the yield $\mean{\underline{s}_1^Z}$ and the upper bound of phase-flip error rate $\mean{\overline{e}_1^{ph}}$ shown in Eq.(\ref{eq:s1LM}) and Eq.(\ref{eq:e1phUM}) respectively. The final key is simply the worst-case result over all possible values for $S_{00}\in [\underline{S}_{00},\overline{S}_{00}]$. $S_{00}$ has a teeny effect on the final key rate as its value is small comparing to the others. With the results of numerical simulations, we conclude that the final key rate only be improved just a little  (less than 0.1 percent) by using these improved methods.

\section{Numerical simulation}\label{sec:Numerical}

\begin{table}
\begin{ruledtabular}
\begin{tabular}{ccccc}
  $p_d$ & $\eta_d$ & $f$ & $\epsilon$ & $e_a$ \\
  \hline
  $1.0\times 10^{-10}$ & $50\%$ & $1.1$ & $1.0\times 10^{-10}$ & $15\%$ \\
\end{tabular}
\caption{\label{tab:parameters} List of experimental parameters used in numerical simulations. $p_d$: the dark count rate, $\eta_d$: the detection efficiency of all detectors, $f$: the error correction inefficiency, $\epsilon$: the security bound considered in the statistical fluctuation analysis, $e_a$: the misalignment error. }
\end{ruledtabular}
\end{table}

\begin{figure}[htb]
  \includegraphics[width=240pt]{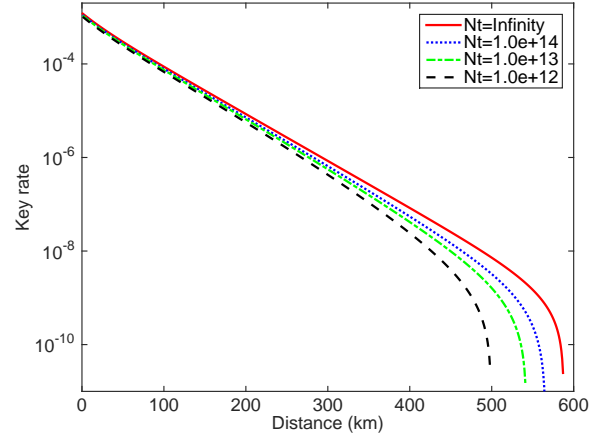}\\
  \caption{(Color online) Optimal secret key rate (per pulse) as a function of the distance with different $N$ by 4-inensity decoy-state method. The asymptotic results are shown in the red solid line. The  blue dotted line, the green dash-dot line and the black dashed line are the results with $N=10^{14}$, $N=10^{13}$ and $N=10^{12}$, respectively.}\label{fig:ROptDiffNt}
\end{figure}

\begin{figure}[htb]
  \includegraphics[width=240pt]{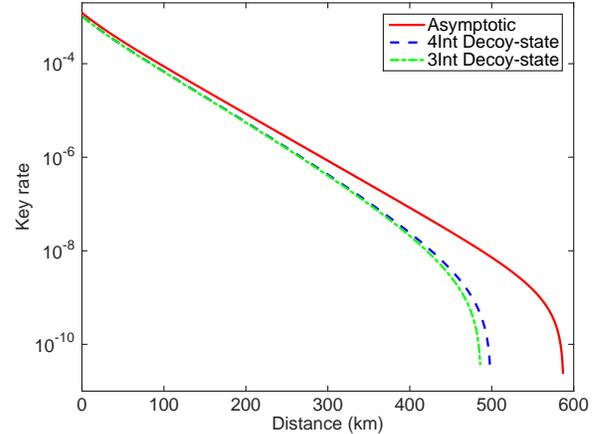}\\
  \caption{(Color online) Optimal secret key rate (per pulse) as a function of the distance by 4-intensity and 3-intensity decoy-state methods with $N=10^{12}$. The asymptotic results with infinite number of pulses are shown in the red solid line. The blue dashed line and the green dash-dot line are the results for 4-intensity and 3-intensity decoy-state methods, respectively.}\label{fig:ROpt3Int4Int12}
\end{figure}

\begin{figure}[htb]
  \includegraphics[width=240pt]{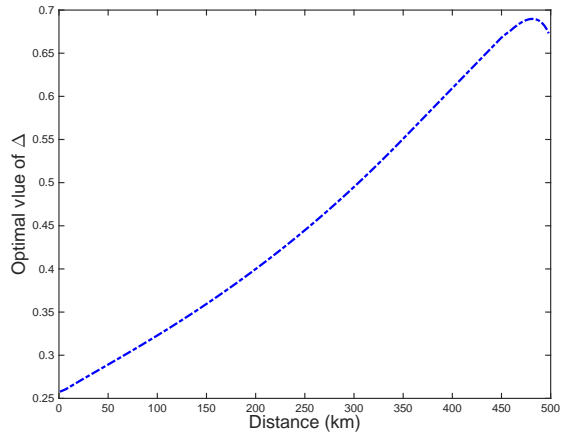}\\
  \caption{(Color online) Optimal value of $\Delta$ corresponding to the optimal secret key rate with $N=10^{12}$.}\label{fig:Delta}
\end{figure}

In this section, we present some results of the numerical simulation. We focus on the symmetric case where the two channel transmissions from Alice to Charlie and from Bob to Charlie are equal. We also assume that Charlie's detectors are identical, i.e., they have the same dark count rates and detection efficiencies, and their detection efficiencies do not depend on the incoming signals. We shall estimate what values would be probably observed in the normal cases by the linear models as previously. The values of the experimental parameters used in the simulations are listed in Table~\ref{tab:parameters}.

We optimize all parameters, $p_X$, $p_1$, $p_2$ $p_z$, $\mu_1$, $\mu_2$, $\mu_z$ and $\Delta$ by the method of full optimization. The results of optimized key rate with different number of pulse pairs by four-inensity decoy-state method are shown in Fig.~\ref{fig:ROptDiffNt}. In it, we use the red solid line to denote the asymptotic results with infinite number of pulses. The optimal key rate with different number of pulse pairs $N=10^{14}$, $N=10^{13}$ and $N=10^{12}$ are shown by the  blue dotted line, the green dash-dot line and the black dashed line respectively. In Fig.~\ref{fig:ROpt3Int4Int12}, we plot the final key rates with the four-intensity decoy-state method and the three-intensity decoy-state method when the number of pulse pairs $N=10^{12}$. We can see that the optimal key rates for the three-intensity decoy-state method is nearly equal to the results for the four-intensity decoy-state method when we are attention to the usability key rate practically (such as $10^{-6}$ per-pulse). The improvement becomes more and more evident when we are regard to the long distance communication. In Fig.~\ref{fig:Delta}, we plot the optimal value of $\Delta$ for different distances with $N=10^{12}$ by four-inensity decoy-state method. 

Also, according to the observed data there~\cite{MDI404km}, we use a linear loss model to estimate the actual loss in the experiment for 404 km of ultralow-loss optical fiber (0.16 dB/km).  Assuming the same device parameter ($p_d=7.2\times 10^{-8}$, $\eta_d=0.5525$, $f=1.16$, $\epsilon=1.0\times 10^{-10}$, $e_a=2\%$ and $N=6.0\times 10^{14}$), we make the optimization by using our sending-or-not protocol with the four-intensity decoy-state method shown above. We obtain a final key rate of 141 bit per second (bps), which is more than $4.4\times 10^{5}$ times higher than the reported experimental result, $3.2\times 10^{-4}$ bps.

\section{Conclusion}\label{sec:Conclusion}
In real setups of QKD, the practical situations with a few different intensities rather than infinite number of different intensities, a phase slice of appropriate size rather than infinitely small size and the statistical fluctuations must be considered. We first present the decoy-state method with a few different intensities and a phase slice of appropriate size. Then we show that the decoy-state method is a highly practical scheme even when the statistical fluctuations are considered. Numerical simulation shows that, the performance of our method is comparable to the asymptotic case for which the key size is large enough. Our results show that practical implementations of the sending-or-not quantum key distribution can be both secure and efficient.

\section*{ACKNOWLEDGMENTS}
We acknowledge the financial support in part by The National Key Research and Development Program of China grant No. 2017YFA0303901; NSFC grant No. 11474182, 11774198 and U1738142; the key Research and Development Plan Project of Shandong Province, grant No. 2015GGX101035; Shandong Peninsula National Innovation Park Development Project; Taishan Scholars of Shandong Province. Z.-W. Yu and X.-L. Hu contributed equally to this work.



\begin{thebibliography}{99}
\bibitem{BB84}
  C.H.~Bennett and G.~Brassard, in {\em Proc.\ of IEEE Int.\ Conf.\ on Computers,
  Systems, and Signal Processing} (IEEE, New York, 1984), pp.~175-179.
\bibitem{GRTZ02}
  N.~Gisin, G.~Ribordy, W.~Tittel, {\em et al.}, Rev. Mod. Phys. {\bf 74}, 145 (2002);
  N. Gisin and R. Thew, Nature Photonics, {\bf 1}, 165 (2006);
  M.~Dusek, N.~L\"utkenhaus, M.~Hendrych, in {\em Progress in Optics VVVX}, edited by E.~Wolf (Elsevier, 2006);
  V. Scarani, H. Bechmann-Pasqunucci, N.J. Cerf, {\em et al.}, Rev. Mod. Phys. {\bf{81}}, 1301 (2009).
\bibitem{ILM}
  H.~Inamori, N.~L\"utkenhaus, and D.~Mayers, European Physical Journal D, {\bf{41}}, 599 (2007), which appeared in the arXiv as quant-ph/0107017;
  D.~Gottesman, H.K.~Lo, N.~L\"{u}tkenhaus, {\em et al.}, Quantum Inf. Comput. {\bf 4}, 325 (2004).

\bibitem{H03}
  W.-Y.~Hwang, Phys. Rev. Lett. {\bf 91}, 057901 (2003).
\bibitem{wang05}
  X.-B.~Wang, Phys. Rev. Lett. {\bf 94}, 230503 (2005).
\bibitem{wang06}
  X.-B.~Wang, Phys. Rev. A {\bf 72}, 012322 (2005).
\bibitem{LMC05}
  H.-K.~Lo, X.~Ma, and K.~Chen, Phys. Rev. Lett. {\bf 94}, 230504 (2005);
  X.~Ma, B. Qi, Y. Zhao, {\em et al.}, Phys. Rev. A {\bf 72}, 012326 (2005).
\bibitem{AYKI}
  Y. Adachi, T. Yamamoto, M. Koashi, {\em et al.}, Phys. Rev. Lett. {\bf 99}, 180503 (2007).
\bibitem{haya}
  M. Hayashi, Phys. Rev. A {\bf{74}}, 022307 (2006); ibid {\bf{76}}, 012329 (2007).
\bibitem{peng}
  D. Rosenberg, J.W. Harrington, P.R. Rice, {\em et al.},  Phys. Rev. Lett. {\bf{98}}, 010503 (2007);
  T. Schmitt-Manderbach, H. Weier, M. R\"{u}rst, {\em et al.}, Phys. Rev. Lett. {\bf{98}}, 010504 (2007);
  C.-Z. Peng, J. Zhang, D. Yang, {\em et al.} Phys. Rev. Lett. {\bf{98}}, 010505 (2007);
  Z.-L. Yuan, A. W. Sharpe, and A. J. Shields, Appl. Phys. Lett. {\bf{90}}, 011118 (2007);
  Y.~Zhao, B. Qi, X. Ma, {\em et al.}, Phys. Rev. Lett. {\bf 96}, 070502 (2006);
  Y. Zhao, B. Qi, X. Ma, {\em et al.}, in {\em Proceedings of IEEE International Symposium on Information Theory, Seattle} (IEEE, New York, 2006), pp.2094--2098.
\bibitem{wangyang}
  X.-B. Wang, C.-Z. Peng, J. Zhang, {\em et al.} Phys. Rev. A {\bf{77}}, 042311 (2008);
  J.-Z. Hu and X.-B. Wang, Phys. Rev. A, {\bf{82}}, 012331(2010).
\bibitem{rep}
  X.-B. Wang, T. Hiroshima, A. Tomita, {\em et al.}, Physics Reports {\bf{448}}, 1 (2007).
\bibitem{njp}
  X.-B. Wang, L. Yang, C.-Z. Peng, {\em et al.}, New J. Phys. {\bf{11}}, 075006 (2009).

\bibitem{PNS1}
  G.~Brassard, N.~L\"utkenhaus, T.~Mor, {\em et al.}, Phys. Rev. Lett. {\bf 85}, 1330 (2000);
  N.~L\"utkenhaus, Phys. Rev. A {\bf 61}, 052304 (2000);
  N.~L\"utkenhaus and M.~Jahma, New J. Phys. {\bf 4}, 44 (2002).
\bibitem{PNS}
  B.~Huttner, N.~Imoto, N.~Gisin, {\em et al.}, Phys. Rev. A {\bf 51}, 1863 (1995);
  H.P.~Yuen, Quantum Semiclassic. Opt. {\bf 8}, 939 (1996).

\bibitem{curty1}
  H.-K. Lo, M. Curty, and B. Qi, Phys. Rev. Lett. {\bf{108}}, 130503 (2012).
\bibitem{wang10}
  X.-B. Wang, Phys. Rev. A {\bf{87}}, 012320 (2013);
  M. Curty, F. Xu, W. Cui {\em et al.}, Nat. Commun. {\bf{5}}, 3732 (2014);
  Z.-W. Yu, Y.-H. Zhou, and X.-B. Wang, Phys. Rev. A {\bf{88}}, 062339 (2013);
  F. Xu, H. Xu, and H.-K. Lo, Phys. Rev. A {\bf{89}}, 052333 (2014);
  Z.-W. Yu, Y.-H. Zhou, and X.-B. Wang, Phys. Rev. A {\bf{91}}, 032318 (2015);
  Y.-H. Zhou, Z.-W. Yu, X.-B. Wang. Phy. Rev. A {\bf{93}}, 042324 (2016).

\bibitem{MDIExps}
  L.C. Comandar, M. Lucamarini, B. Fr\"ohlich, {\em et al.}, Nature Photonics {\bf{10}}, 312 (2016);
  C. Wang,~Z.-Q. Yin,~S. wang,~W. Chen,~G.-C. Guo,~Z.-F. Han, Optica, {\bf{4}}, 1016 (2017).

\bibitem{MDI404km}
  H.-L Yin, T.-Y Chen, Z.-W Yu, {\em et al.}, Phy. Rev. Lett. {\bf{117}}, 190501 (2016);

\bibitem{ind3}
  S.L. Braunstein and S. Pirandola, Phys. Rev. Lett. {\bf{108}}, 130502 (2012).
\bibitem{ind2}
  H.-K. Lo, M. Curty, and B. Qi, Phys. Rev. Lett., {\bf{108}}, 130503 (2012);
  K. Tamaki, H.-K. Lo, C.-H. F. Fung, {\em et al.}, Phys. Rev. A, {\bf{85}}, 042307 (2012).

\bibitem{nature18}
  M. Lucamarini ,Z.L. Yuan, J.F. Dynes, \& A.J. Shields, Nature 557, pages 400¨C403 (2018)
\bibitem{mxf}
  Xiongfeng Ma, Pei Zeng, Hongyi Zhou, arXiv:1805.05538, 2018, {\em Phase-matching quantum key distribution}
\bibitem{star}
  K. Tamaki, H.-K. Lo, W. Wang, and  M. Lucamarini, arXiv:1805.05511, 2018, {\em Information theoretic security of quantum key distribution overcoming the repeaterless secret key capacity bound.}
\bibitem{wxb}
  Xiang-Bin Wang, Zong-Wen Yu, and Xiao-long Hu, arXiv:1805.09222, 2018, {\em Sending or not sending: twin-field quantum key distribution with large misalignment error}

\end{thebibliography}
\end{document}